\documentclass[twocolumn]{aastex631}

\setcitestyle{authoryear,open={(},close={)}}

\def\nH    {n$_{\scriptscriptstyle H}$}
\def\NH    {$\rm N_{\scriptscriptstyle HI}$}
\def\Msun  {${\rm M}_\odot$}
\def\deg   {$^\circ$}
\def\arcmin {$^\prime$}

\def\kms   {\ km s$^{-1}$}
\def\cm   {\ cm$^{-2}$}
\def\Mhi   {M$_{\rm HI}$}

\def\Msun   {M$_{\odot}$}

\shorttitle{3D HI Filaments}
\shortauthors{Putman et al.}

\begin{document}

\title{A Catalog of Galactic Atomic Hydrogen Position-Position-Velocity Filaments}

\author[0000-0002-1129-1873]{M. E. Putman}
\affiliation{Department of Astronomy, Columbia University, New York, NY 10027, USA}
\author[0000-0001-9654-5889]{D.A. Kim}
\affiliation{Department of Astronomy, Columbia University, New York, NY 10027, USA}
\author[0000-0002-7633-3376]{S. E. Clark}
\affiliation{Department of Physics, Stanford University, Stanford, CA 94305, USA}
\affiliation{Kavli Institute for Particle Astrophysics \& Cosmology, P.O. Box 2450, Stanford University, Stanford, CA 94305, USA}
\author{L. Li}
\author[0009-0002-1128-2341]{C. Holm-Hansen}
\affiliation{Department of Astronomy, Columbia University, New York, NY 10027, USA}
\affiliation{Department of Astronomy, University of Michigan, Ann Arbor, MI 48109, USA}
\author[0000-0003-4797-7030]{J.E.G. Peek}
\affiliation{Space Telescope Science Institute, 3700 San Martin Drive, Baltimore, MD 21218, USA}

\correspondingauthor{M.E. Putman}
\email{mep2157@columbia.edu}

\begin{abstract}
We present a catalog of 3D Galactic HI filaments over $1/3$ of the sky using Galactic Arecibo L-band Feed Array HI (GALFA-HI) data.   The 3D filaments are defined to be linear HI features that are continuous in position-position-velocity (PPV) and are found with {\it fil3d}, an algorithm that expands on the 2D {\it FilFinder}. 
The catalog contains 3333 HI filaments between $\pm50$\kms~at a range of Galactic positions. 1542 of the PPV filaments are identified as local at the distance of the wall of the Local Bubble, and 209 are likely at the disk-halo interface of our Galaxy.  The catalog and properties of the PPV filaments are obtained after an unsharp mask (USM) is applied to the data.  The widths of the filaments are consistently $\sim12$\arcmin~ (0.34 pc at 100 pc), and constrained by the 4\arcmin~resolution.
The local filaments have median properties of N$_{\rm HI(med)}$(USM) = $6 \times 10^{18}$\cm, \Mhi = 0.17 \Msun, FWHM = 3.2\kms, and length of 6.4 pc.  The disk-halo population has similar column densities, but the median FWHM = 7.7\kms, consistent with them being higher z-height, warmer structures.  The L $\propto$ M$^{0.5}$ relationship found for the HI filaments and their bundling on the sky are consistent with a hierarchical structure, and is likely related to turbulence playing a role in their formation.

\end{abstract}


\section{Introduction}
\label{intro}

Gaseous filamentary structure permeates the Galactic interstellar medium (ISM).   The filaments' wide range of densities, environments, and compositions indicate they represent physical mechanisms in the ISM that transcend any one scale or context.    
The filaments have primarily been catalogued and characterized in the dense, dusty and/or molecular ISM \citep[e.g.,][]{Arzoumanian:2011wu,Palmeirim:2013td,Malinen:2016}.   Atomic hydrogen (HI) is the dominant gaseous component of the ISM and the transition point between the ionized baryons that dominate the Universe and star-forming molecular structures.  Filaments have been found across the sky in HI surveys \citep[e.g.][hereafter C14]{soler22,kim23,kalberla16,clark14}.    Determining the properties of this population paves the way towards understanding the physical mechanisms creating them.

The long axis of the HI filaments (also referred to as HI fibers) are aligned with the Galactic magnetic field at high latitude, but show variations at lower Galactic latitudes \citep[C14,][]{clark15,soler20,soler22,kim23,holm24}.  The alignment at high latitude is stronger with the 4\arcmin~GALFA-HI (Galactic Arecibo L-band Feed Array HI) data compared to the 16\arcmin~HI4PI data (C14), and this indicates the filaments are better characterized with higher resolution data.  The diffuse atomic filaments are not self-gravitating, and alignment with the magnetic field is not expected when the filaments are self-gravitating \citep{Soler:2013dh, FalcetaGoncalves:2008wo}. Indeed, observations of denser structures have often found anti-alignment with the Galactic magnetic field \citep{Collaboration:2016XXXV,Collaboration:2016, Palmeirim:2013td,stephens22}. 

Theoretical work on the atomic filaments remains limited, partially due to their observational properties not being well constrained.  \cite{Hennebelle:2013ws}  and \cite{Inoue:2016wb}  find that the atomic filaments can align with the magnetic field if the turbulent shear strain is aligned with the magnetic field.  This could be consistent with the bulk of the filaments overhead being related to compression along the Local Bubble \citep{ntormousi11,alves18}.  The flow of lower density filaments into molecular self-gravitating filaments has also been predicted by simulations \citep{soler17}, and is hinted at in molecular gas observations \citep{kirk13,Peretto:2012bu, Palmeirim:2013td}. Further observational constraints on the HI filaments can guide future models and reveal the nature of the interaction between the magnetic field and the ISM. 

The HI filaments have only recently begun to have their properties more broadly characterized. The alignment of the HI filaments with the magnetic field was first quantified by C14 with an algorithm that identifies linear structures in fixed velocity width images. This work examined one of the magnetically aligned filaments and found a FWHM linewidth of 3.4\kms~and a column density of $5.3 \times 10^{18}$ cm$^{-2}$. 
\cite{kalberla16} examined filamentary structure found in the lower resolution HI4PI data by completing an unsharp mask and examining the properties of features at $|b| >$ 20\deg~that appeared filamentary.  They found typical column densities of $10^{19}$ cm$^{-2}$ and a similar typical velocity width to C14 for their filaments.  
Both C14 and K16 found the filaments tend to be in bundles at similar velocities, and neither developed a method to successfully catalog and extract the properties of individual filaments.  \cite{soler22} did a similar analysis to K16 of bright filamentary structures in the Galactic Plane and found the orientation of the structures to change from being perpendicular to the plane in the inner Galaxy, to parallel to the plane in the outer regions.

In this work, we assume that the linear HI structures are cylindrical filaments, with one long axis and two short axes. It has been shown that the filaments under examination are indeed physical structures. They have distinct far-IR emission per 21-cm emission \citep{CPMD} and distinct absorption features \citep{CP19} from the surrounding material, and thus are not caused by ``velocity caustics''.  It has not yet been fully proven that they are not caused (at least in part) by projection effects, in which edge-on sheet-like structures appear linear in our observations.   Given the bundling of the filaments and their narrow velocity widths, it seems unlikely to be a reasonable physical model for the linear structures.  

In this paper, we catalog and analyze the properties of 3D (position-position-velocity) HI filaments found across the 13,000 deg$^2$ of sky observed in GALFA-HI \cite{peek18}.  This 4\arcmin~and 0.18\kms~resolution survey 
is ideal for extracting the properties of the fibrous structure of our Galaxy using a newly developed tool that joins 2D FilFinder \citep{Koch:2015dc} identified filaments in velocity space.   In \S2 we briefly describe the GALFA-HI data and in \S3 we discuss our method of extracting three-dimensional filaments with the algorithm we call {\it fil3d}.   We present the catalog of HI filaments, their distance estimates, and define a local and disk-halo population in \S4.   The filament catalog values, and derived physical properties, are presented in \S5. Finally, we discuss and conclude this study of the filamentary HI sky in \S6 and \S7.


\section{The GALFA-HI Data}
\label{sect:galfa}

The data used for this HI filament study are from the GALFA-HI survey DR2 \citep{peek18}.  GALFA-HI is a relatively high resolution ($\sim4'$), large area (13,000 deg$^2$), high spectral resolution (0.18\kms), wide bandwidth ($\pm$650\kms) HI survey conducted with the multibeam receiver on the Arecibo radio telescope.  We analyze the entire area covered by the survey (Declinations = $-1$\deg to $+37$\deg~and all Right Ascensions; see Figure~\ref{sky}), but we limit our velocity range to $+/- 50$\kms~and use the cubes smoothed to 0.754\kms~per channel.  This choice of velocity range was made early in the study given the computing resources and in order to focus on Galactic emission.  In the future, a larger velocity range would be of interest.  In terms of Galactic coordinates, the survey covers the complete range of absolute Galactic latitudes and passes through the Galactic Plane at approximately longitude 190\deg~and 45\deg \citep[see figure 1 in][]{peek18}.  
GALFA-HI DR2 data have a typical $5\sigma$ column density sensitivity of $\sim10^{18}$ cm$^{-2}$ per 1~\kms.  The fibrous structure within the GALFA-HI DR2 data were presented for binned 3\kms~velocity slices in \cite{peek18} using the Rolling Hough Transform (RHT) method pioneered in C14. The fact that the RHT was not able to identify the filaments as true 3D structures, nor subsequently catalog their spatial and kinematic properties, was the main motivation for this study.

\section{Extracting 3D Filaments}
\label{methods}

Most filament finding algorithms require that 3D data is binned in velocity space to create 2D maps before they are analyzed.  The goal of this work was to create a method of cataloging the filaments in three dimensions without initial velocity binning.  This makes it less likely the filaments are chance projections created by the binning \citep[e.g.,][]{Panopoulou:2014kd}, and allows the kinematic properties to be derived in addition to the spatial properties.  To achieve this goal we adapted the two-dimensional {\it FilFinder} \citep{Koch:2015dc} to work on three dimensional data and refer to the algorithm as {\it fil3d}.  
We describe the logic behind the choice of starting algorithm, our adaptation of it to three dimensions, and the parameters used for the GALFA-HI data below.  See \cite{Koch:2015dc} for the details on the 2D {\it FilFinder} code.

\subsection{Starting with the 2D {\it FilFinder}}
\label{sect:start}
{\it FilFinder} is particularly appealing for the study of low density HI filaments because it is able to identify filaments over a wide range of brightness. 
A number of methods for characterizing filamentary structure have been developed, and we have refined our current approach after considering several alternatives. For instance, {\it DisPerSE} \citep{Sousbie:2011ft} is designed to highlight filaments that link overdensities, and is optimized for the cosmic web and structures caused by gravitational collapse. \cite{Koch:2015dc} did a comparison between {\it FilFinder} and {\it DisPerSE} and found that while the algorithms recover similar structures in dense regions, {\it DisPerSE} was unable to reliably recover the faint structures that {\it FilFinder} was able to identify.

Some analyses have used eigenvectors of the Hessian matrix to identify regions of high local curvature. Individual filaments can then be identified by imposing additional criteria to segment the data. Hessian-based approaches have been applied to the cosmic web
\citep{bond10}, dust emission \citep{Collaboration:2016}, and diffuse HI data \citep{kalberla21, Cukierman:2023, halal24}. 
\cite{soler20} found consistent results with the Hessian and {\it FilFinder} for bright filaments in the Galactic Plane. The algorithm {\it getfilaments} \citep{Menshchikov:2014we} was specifically designed to work on \textit{Herschel} data in the presence of contaminating bright sources and is not the best choice given our goals. 

The adaptive thresholding in {\it FilFinder} allows us to identify HI features over a large dynamic range. 
{\it FilFinder} first prepares a mask for the data by flattening and smoothing the image and applying an adaptive threshold where the central pixel of a patch must be greater than the median of the neighborhood.   Filaments are identified within the mask with a medial axis transform that puts objects into one pixel wide skeletons maximized in one direction.  The final step of {\it FilFinder} is to prune the skeletons by removing small offshoots from the main long axis of a filament.  

\subsection{{\it fil3d}:  The 3D Filament Finder}
\begin{figure*}[t]
\includegraphics[scale=0.45]{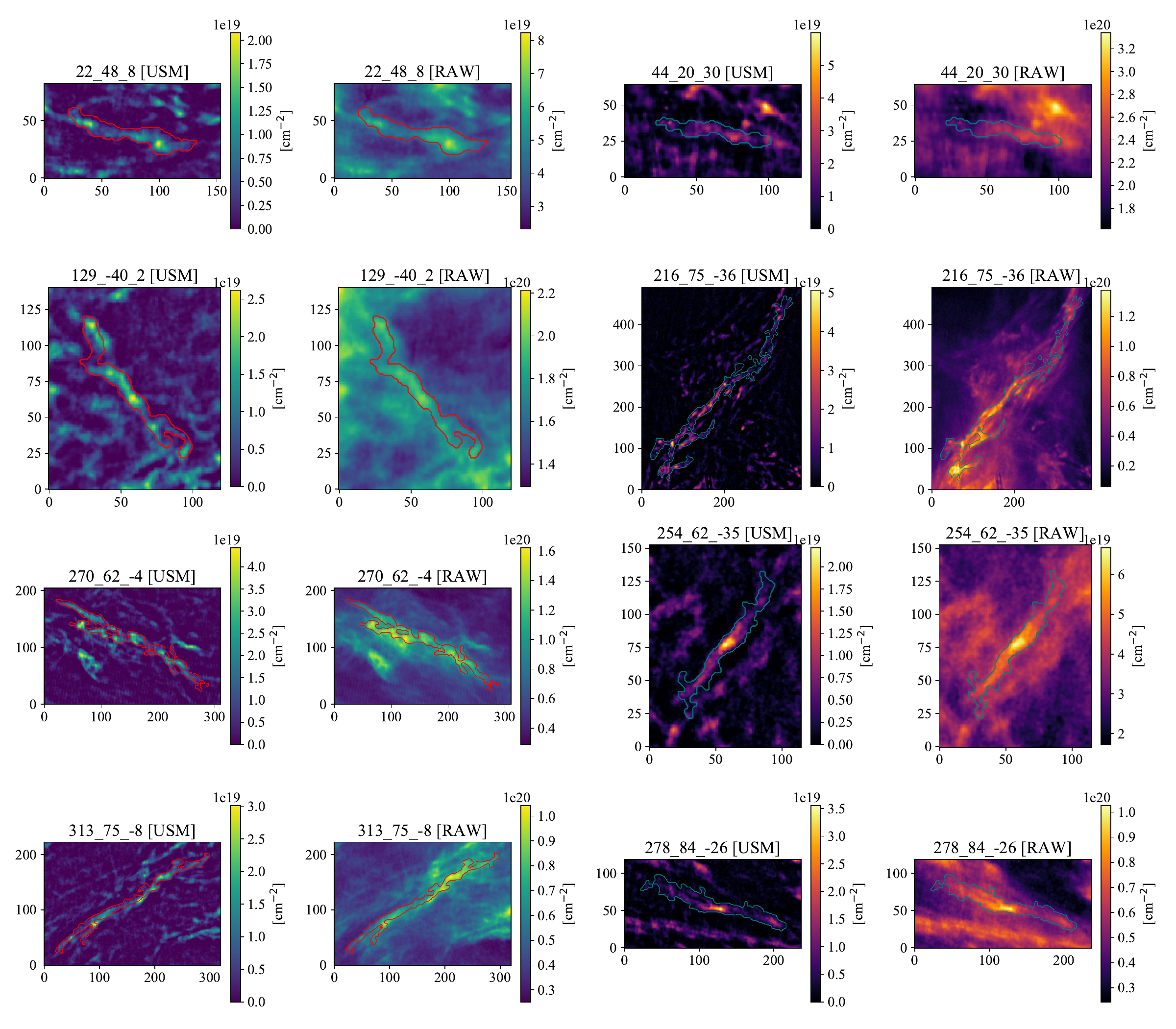}
\caption{Moment 0 maps of example 3D HI filaments from the local (left two columns) and disk-halo (right two columns) populations.  The red or green contour lines are the filament masks as identified by {\it fil3d} and the background is the USM GALFA-HI data (labelled ``USM") for the first column of each population and the original raw data for the second column (labelled ``RAW"), integrated over the FWHM velocity width. The axes are in pixels which are 1\arcmin.  It is evident that the USM successfully extracts the filament, but also removes a diffuse component that affects the derived column densities and widths.  }
\label{masks}
\end{figure*}

Our three-dimensional filament finder utilizes {\it FilFinder} with the steps outlined here.   The specific parameters used for this 3D filament catalog are also noted. In addition to this work, we have used {\it fil3d} to catalog GALFA-HI filaments in higher velocity resolution data at high Galactic latitude \citep{kim23} and in a small region towards a high-velocity cloud \citep{holm24}.  The {\it fil3d} code is distributed via GitHub at \url{https://github.com/a-dykim/fil3d}.
\begin{enumerate}
    \item For each GALFA-HI 0.754\kms~spaced velocity channel an unsharp mask (USM) with a Gaussian FWHM of 30\arcmin~is applied to remove the large scale diffuse Galactic emission.  The scale of the USM was chosen to optimize the removal of diffuse, large-scale Galactic emission in the GALFA-HI data.  Different values are likely suitable for other datasets.  
\item  {\it FilFinder} is run on each velocity channel and objects are identified by their extremal RA and DEC positions.  We adopt a scale width parameter of 0.1 pc at a distance of 100 pc, which is based on 0.1 pc being our resolution at this distance.  The scale width is used to set the smoothing scale (scale width/2) and the adaptive thresholding scale (scale width$\times2$) within {\it FilFinder}.  We linked the size threshold to the scale width parameter as (scale width$\times 2$)$^2 \times 8$, which corresponds to an elongated filament structure.  Using lower values for the size threshold results in a large number of additional filaments, but the vast majority of the additional small structures are then filtered out in the velocity width and aspect ratio cuts outlined in \S\ref{sect:cuts}.

 \item The areas of the {\it FilFinder} objects in adjacent channels are compared, and if there is a $>85$\% overlap, they are merged to be part of the same structure.  This comparison to objects in adjacent channels is continued until there is no longer a $>85$\% area overlap between objects.  At this point the 3D structure is defined and the overlapping channels merged to form the 3D filament mask (see Figure~\ref{masks}).  The overlap fraction was tested by visually inspecting the results with different values to check for filaments with additions that did not belong, or evidence for the truncation of some filaments.  {\it fil3d} does not discard objects that are not found to overlap with a filament in another channel (i.e., single-channel objects remain in the catalog at this stage).  The overlap fraction can be adjusted to suit the dataset and environment for future filament catalogs \cite[see further discussion in][]{holm24}.   For instance, the overlap fraction may need to be smaller for datasets where the filaments are expected to have a large velocity gradient.

\end{enumerate}


\subsection{Selection cuts for the 3D filament catalog}
\label{sect:cuts}

After running {\it fil3d} on the GALFA-HI data we applied several selection cuts to ensure we had a well-defined set of 3D filaments. 
\begin{enumerate}
    \item The filaments were required to extend over two or more channels ($\Delta v \geq 1.47$\kms) to select only the 3D filamentary structures (i.e., single-channel objects were discarded).  
    
    \item 3D structures with aspect ratios of at least 1:4 were selected to ensure the objects are elongated. The aspect ratio was calculated from the ratio of the short to long axis of the 3D filament mask.  The initial list had 63,270 3D objects, and over half of the objects had a 1:1 aspect ratio.  We chose 1:4 over other ratios as this resulted in a visually clean set of filamentary structures and the properties of the selected filaments did not significantly change with different ratio cuts.  There were 3671 filaments with a 1:4 or larger range aspect ratio.  


\item The next selection cut ensures the filament is at the velocity peak of the HI emission within the 3D filament mask.  We use the USM data to estimate the velocity full width at half maximum (FWHM) of the filament using the method developed by \cite{kim23}. 
We use the spatial area of the filament mask, but extend to channels beyond the velocity range of the mask to obtain a median intensity line profile for the filament.   We then fit a Gaussian to this line profile and ensure that the {\it fil3d} identified channels are within 1$\sigma$ of the center of the Gaussian.  This check removes $\sim250$ filaments from the 3671.

  \item Finally, we removed filaments that visually ran along the declination edge of the GALFA-HI observed region and those objects that were artifacts from the basketweave scans that had not yet been excluded by the other cuts \citep[see][for more details on the data properties]{peek18}.  This removed $\sim92$ additional filaments.
  
\end{enumerate}
  

The final catalog of continuous linear structures in position-position-velocity space between $\pm50$\kms~over the $\sim$13,000 deg$^2$ GALFA-HI region has 3333 filaments.  The spatial mask for each filament, and its average velocity, are shown in Figure~\ref{sky}.    

\begin{deluxetable*}{lccccccccccccccc}
\label{tab:cat}
\tablecaption{The 3D HI Filament Catalog}
\tabletypesize{\scriptsize}
\tablehead{
\colhead{ID} & \colhead{$l$}  & \colhead{$b$}  & \colhead{$v_{\it LSR}$} & \colhead{$\Delta v_{\it fil3d}$} &  \colhead{FWHM} & \colhead{len} & \colhead{N$_{\rm HI(tot)}$}  & \colhead{N$_{\rm HI(med)}$} & \colhead{N$_{\rm HI(med)}$} & \colhead{r$_{usm}$} & \colhead{r$_{raw}$} & \colhead{d$_{kin}$} & \colhead{d$_{\rm LB}$} & \colhead{$\theta$} \\
\colhead{} & \colhead{\deg} & \colhead{\deg} & \colhead{\kms} & \colhead{\kms} & \colhead{\kms} & \colhead{\arcmin} & \colhead{(USM)cm$^{-2}$}  & \colhead{(USM)cm$^{-2}$} & \colhead{(raw)cm$^{-2}$} & \colhead{\arcmin} & \colhead{\arcmin} & \colhead{pc} & \colhead{pc} & \colhead{\deg} 
}
\startdata
31\_21\_-9 & 31.93 & 21.69  & -9.6 & 2.2 & 4.0 & 127 & 5.6E+21 &  4.0E+18 & 4.9E+19 & 6.0 & 10.3 & 0 & 80 & 14 \\ 
349\_73\_-1 & 349.98 & 73.76  & -1.5 & 1.5 & 7.0 & 124 & 1.3E+22 &  9.4E+18 & 1.2E+20 & 5.6 & 11.4 & 0 & 155 & 9 \\ 
99\_-23\_5 & 99.01 & -23.82  & 5.2& 2.2 & 2.9 & 141 & 7.2E+21 &  4.4E+18 & 8.8E+19 & 5.1 & 10.5 & 1154 & 100 & 27 \\ 
329\_69\_-6 & 329.18 & 69.82  & -6.6 & 1.5 & 5.4 & 151 & 9.3E+21 &  5.2E+18 & 9.7E+19 & 6.2 & 11.3 & 0 & 145 & -20 \\
167\_-9\_9	& 167.87 &	-9.13 & 9.6 &	3.7 &	3.2	& 409	& 1.1E+23	& 1.1E+19	& 5.1E+20	& 7.6	& 15.2 &	1166	& 115 &	-31 \\
\enddata
\tablecomments{See \S\ref{sect:subprops} for a full description of each column.  The complete catalog is available electronically.}
 
\end{deluxetable*}

\section{HI Filament Catalog}
\label{sect:props}

The properties of the 3333 3D HI filaments are presented in Table~\ref{tab:cat}, with the full table available electronically.  This section describes each column of the table, how distances to the filaments are derived, and how some filaments are grouped into 'local' and 'disk-halo' populations.

\subsection{Catalog Description}
\label{sect:subprops}
The columns of the table can be described as follows:
Column 1 (ID) -- The 3D filament ID, which is the Galactic latitude and longitude and average velocity (LSR) from the following columns truncated to whole numbers.

Columns 2 \& 3 ($l,b$) -- The center of mass of the filament in Galactic coordinates. This is calculated using the filament mask identified by {\it fil3d}. Note that in cases where the filament is significantly curved or has a complex structure, this position can be outside of the contours of the filament.

Column 4 ($v_{LSR}$) -- The average LSR velocity of the filament.  This velocity is from the {\it fil3d} moment 1 map and would be similar if obtained from the channels included in the Gaussian fit since the {\it fil3d}-identified velocities need to be within 1$\sigma$ of the Gaussian peak. 

Column 5 ($\Delta v_{fil3d}$) -- The velocity extent of the filament found by {\it fil3d} to have a continuous, significantly overlapping filamentary structure.  The velocity span is in multiples of 0.754\kms~due to the channel spacing of the data.

Column 6 (FWHM) -- The FWHM of the Gaussian fit to the line profile obtained using the median intensity in each channel within the spatial filament mask. This is from the USM data and explained further in the third step of \S\ref{sect:cuts}.  

Column 7 (len) -- The length of the filament in arcminutes obtained by projecting the long axis of the filament mask to a flat x-axis and measuring the number of 1\arcmin~pixels across the filament.   

Column 8 (N$_{\rm HI(tot)}$(USM)) -- The total column density within the mask of the filament over the FWHM velocity range using the USM data.  This value will have some of the diffuse emission associated with the filament removed.


Column 9 (N$_{\rm HI(med)}$(USM))-- The median column density within the mask of the filament over the FWHM velocity range evaluated with USM data.  

Column 10 (N$_{\rm HI(med)}$(raw)) -- The median column density, N$_{\rm HI(med)}$(raw), within the mask of the filament over the FWHM velocity range using the raw data. This density is an overestimate with its inclusion of diffuse Galactic emission.  

Column 11 (r$_{usm}$) -- The radius of the filament in arcminutes derived from the USM data by running {\it FilFinder} on the moment 0 maps created using the 3D filament masks.  The radius is derived by fitting a Gaussian profile along the skeleton of the filament.
See \cite{Koch:2015dc} for more details.   This value represents one half of the filament and should be doubled for the full width of the filament (as is done throughout the rest of the paper).  Only 0.6\% of the filaments have a failed width fit to the USM data as represented by a 0 or an unusually large value in the catalog.

Column 12 (r$_{raw}$) -- The radius of the filament in arcminutes similar to the previous column except using the raw GALFA-HI data.
Without the USM, the diffuse Galactic emission often confuses the fit, making this value less reliable.  Approximately 15\% of the filament fits to the raw data are complete failures as represented by a 0 or a large value in the catalog.

Column 13 ($d_{\rm kin}$)-- The kinematic distance derived for the filament (described below in \S\ref{sect:dist}).  Filaments that do not have a position and velocity that enables a kinematic distance are placed at 0 or 10,000 pc.

Column 14 ($d_{\rm LB}$) --  The distance derived for the filament with the assumption that it is at the edge of the Local Bubble \citep[see \S\ref{sect:dist};][]{capitanio17}. 
 This is a lower limit on the distance and is only adopted for the local filaments defined in \S\ref{sect:pops}.

Column 15 ($\theta$) -- The orientation angle of the filament relative to 0\deg\ declination.  A negative value is counter-clockwise to this reference and a positive value is clockwise. The filaments are not long enough for the curvature of the sky to have a large effect on this value.  


\begin{figure*}[t]
\hspace{-6cm}
\includegraphics[scale=0.5]{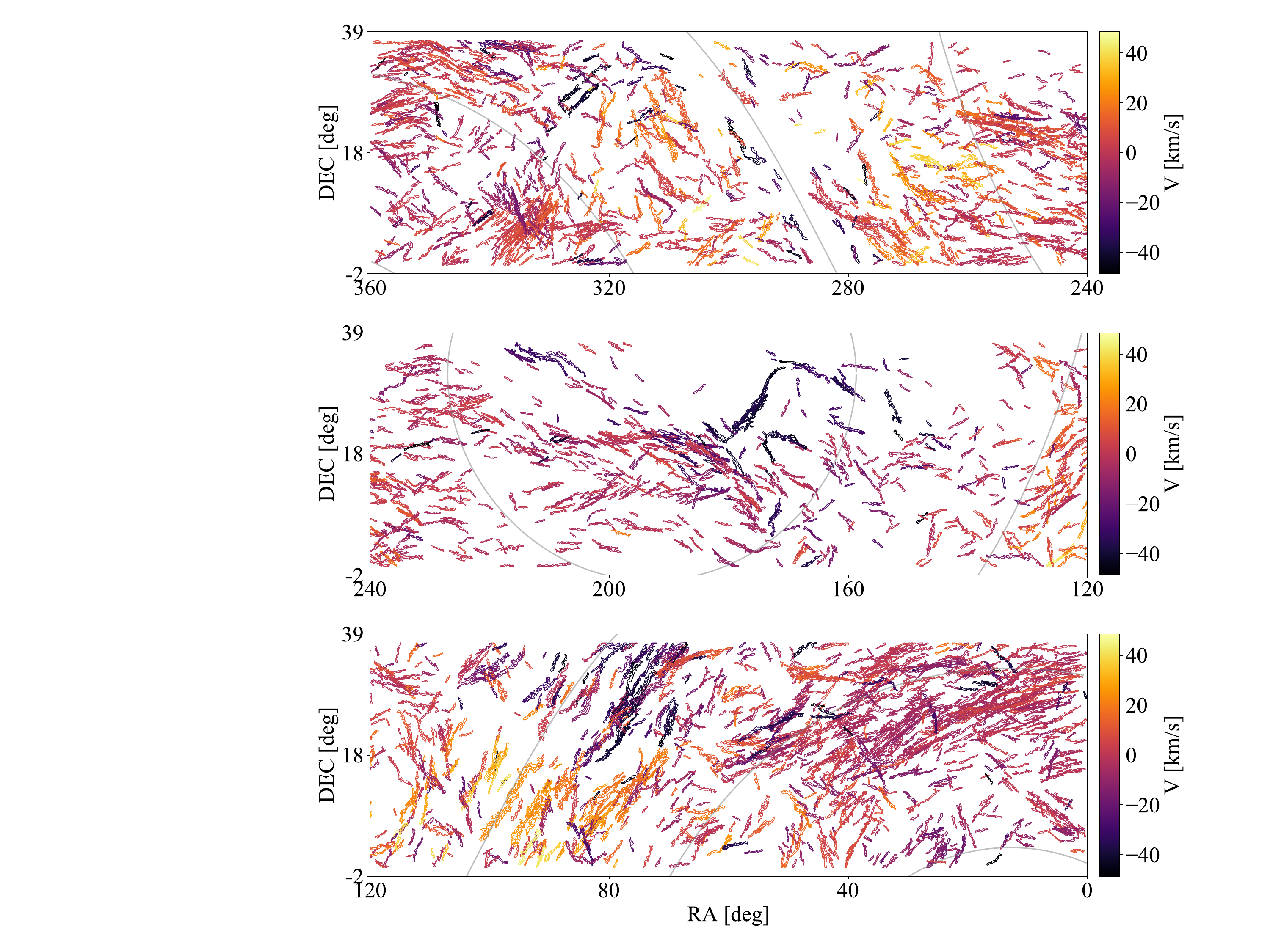}
\caption{The distribution of the 3D filament catalog on the GALFA-HI survey area with the color showing the average velocity of each filament in the LSR frame.  The coordinates are noted in RA (x-axis) and Dec (y-axis) in degrees and the grey lines on the background represent different Galactic latitudes in 30\deg~increments (top: -30\deg, 0\deg, 30\deg (inner Galaxy); middle: 60\deg, 30\deg; bottom: 0\deg, -30\deg, -60\deg (outer Galaxy)). }
\label{sky}
\end{figure*}

\subsection{Filament Distances}
\label{sect:dist}
To derive many physical properties of the filaments, distances must be assigned.  We do not have a precise method of assigning a distance to a filament, but we have developed two methods of estimating the distance to a filament. The first method assigns a lower limit on the distance based on the distance to the wall of the Local Bubble ($d_{\rm LB}$) in the filament direction \citep{capitanio17,Lallement:2014wr}.  This distance estimate is for those filaments near 0\kms~(see \S~\ref{sect:pops}), but we include the derived value for all filaments.
A fiducial value of E(B-V) = 0.015 is used for the reddening at the near wall of the local cavity. Using the \citet{capitanio17} map,
we obtain the distance at which E(B-V) first reaches 0.015 for the $(l, b)$ catalog position of each filament.  The associated distance uncertainty for this edge is typically between 15-30 pc. For several sky positions this reddening is not reached out to the farthest distance in the catalog, and for these we take the maximum distance in the map as a lower limit on the distance to the filaments.   

The second distance estimate method assigns a kinematic distance ($d_{\rm kin}$) to each filament, as
rotation curves are close to flat within the nearest kpc \citep{Tchernyshyov:2017hw, duval09}.  There is a direct
relationship between distance and velocity in the outer Galaxy. In the inner Galaxy the standard
distance ambiguity effects are not a problem for filaments with $|b| > 10$\deg, as assuming a “far side”
distance would place the filaments unphysically far above the Galactic Plane.   As can be seen from Figure~\ref{sky}, the majority of our filaments are not clustered around the plane in the inner Galaxy as we have not optimized the finder to deal with the density of structures there. In regions like this, kinematic distances should be used with caution given the likely uncertainties \citep[e.g.,][]{hunter24, soler25}.  The distribution of kinematic distances is limited by the $\pm 50$\kms~range chosen for this study.

 \subsection{Local and Disk-Halo Populations}
\label{sect:pops}
For some filaments we use the positions, velocities and distances to identify them as being near the edge of the Local Bubble (local), or in the region between the disk and the halo (disk-halo).     

To isolate the filaments that are most likely local, we select filaments with $|v_{\rm LSR}| < 10$\kms \citep{dickey90, soler25, soding25}, and exclude all filaments towards the outer Galaxy (90\deg $< l <$ 270\deg) with $b<30$\deg.  The latter cut is required because gas at low velocities in this direction can be at a wide range of distances.  With this definition, there are 1542 local filaments, and the edge of the Local Bubble distances are adopted for these filaments.  

\begin{figure}[t]
\includegraphics[width=0.52\textwidth]{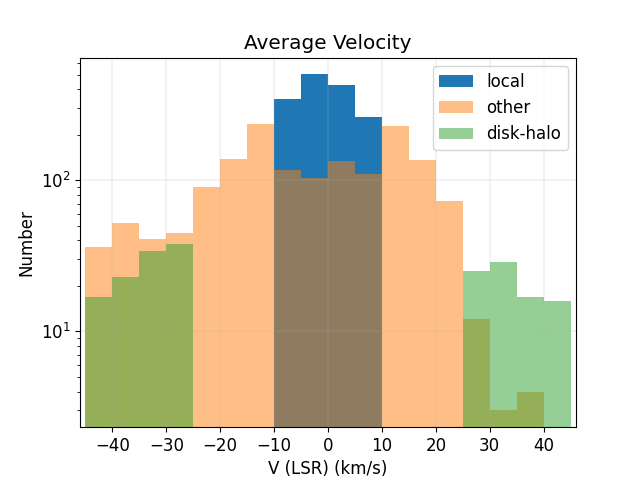}
\caption{Histogram of the average velocities of the local, disk-halo, and other 3D filaments with the y-axis the number of filaments on a log scale.   The `other' population may have local or disk-halo filaments within it, but their locations are less certain.}
\label{velocity}
\end{figure}

For the remaining 1791 filaments, the kinematic distance is favored.
From this group, we identify a disk-halo population that is defined to have absolute velocities $>25$\kms~and have either a z-height $>200$~pc (108 filaments) or a failed kinematic distance for $b >$~20\deg (101 filaments). The median distance of the 108 disk-halo filaments with kinematic distances is 5.4 kpc and the median z-height is 1.1 kpc.  
The velocity distributions of the populations of filaments are shown in Figure~\ref{velocity}, with the local and disk-halo populations noted.  We note that the population labeled ``other" likely contains a mixture of local, disk and disk-halo filaments, but they cannot be as clearly defined.  The median properties of the populations of filaments are noted in Table~\ref{tab:med}.

\begin{deluxetable*}{lcccccc}
\tablecaption{Median properties of the 3D HI filaments}
\tablehead{
\colhead{Population} & \colhead{Total} & \colhead{v$_{LSR}$}  & \colhead{FWHM} & \colhead{N$_{\rm HI(med)}$} & \colhead{N$_{\rm HI(tot)}$}  & \colhead{d}\\
\colhead{} & \colhead{No.} & \colhead{\kms} & \colhead{\kms} & \colhead{(USM)~cm$^{-2}$} & \colhead{(USM)~cm$^{-2}$}  & \colhead{pc}
}
\startdata
       Local$^{a}$  & 1542 & -0.8  & 3.2  & 5.9E+18 & 1.4E+22 & 130\\
       Disk-halo$^{b}$ & 209 & -26.5  & 7.7 & 6.3E+18 & 1.8E+22  & 5383 \\
        Other$^{c}$ &  1582 &  -4.0  & 4.6 & 7.7E+18 & 1.9E+22 & 2329 
\enddata
\tablenotetext{a}{This includes filaments with $|v_{LSR}| < 10$\kms, but excludes filaments with 90\deg $< l < $270\deg~and $b < 30$\deg~due to ambiguous distances in this direction. The median distance is from $d_{\rm LB}$ (see \S\ref{sect:dist}).}
\tablenotetext{b}{This is all of the filaments with  $|v_{LSR}| > 25$\kms~and either a z-height $> 200$ pc or a failed kinematic distance for those at $b>20$\deg.  The median distance is for the 108 filaments with a $d_{\rm kin}$.}
\tablenotetext{c}{This is all filaments that are uncategorized (not in the local or disk-halo sample).  The median distance is for the 636 of these filaments with a $d_{\rm kin}$.}
\label{tab:med}
\end{deluxetable*}

\section{Results}
\subsection{Sky Distribution}
\label{sect:sky}
Figure~\ref{sky} shows the distribution of filaments on the sky, with each filament color coded by its average velocity.  There are filaments near 0\kms~across the majority of the sky, with a clear deficit close to the Galactic Plane.  The lack of 0\kms~filaments at low latitude (in particular the inner Galactic Plane in the top panel) does not indicate they are not there, but rather that it is a complex region where filaments are more difficult to recover.  The top panel has $b=0$\deg\ as the middle line and is centered at $l \sim 50$\deg\ and so the negative velocity filaments shown are consistent with Galactic rotation in this direction.  Likewise, the outer 
Galactic Plane in the bottom panel is shown by the first line in Figure~\ref{sky} at $l \sim 190$\deg, and many of the positive and negative velocity filaments are consistent with Galactic rotation.

Beyond the region close to the Galactic Plane, there are filaments at low velocities that are likely close to us (within 200 kpc) and filaments that have velocities that place them beyond what is expected from Galactic rotation.  This latter population of filaments is likely linked to intermediate velocity clouds (IVCs).  A 360 deg$^2$ high latitude region in the center of the middle panel of Figure~\ref{sky} was further studied by \cite{kim23} with {\it fil3d} using GALFA-HI data with 0.18 \kms~channel spacing and a similar distribution of low velocity and intermediate velocity filaments was identified.


The filaments tend to be grouped into bundles with similar velocities and similar alignment.  This is consistent with the overall distribution of fibers found with the RHT in P18.\footnote{We note that there is an error in the figure caption for the filament map in \cite{peek18} in that the velocity color scale should be reversed.}   We have not done a direct correlation with the RHT results for this paper, but there is a visual similarity to the filament groupings.  It is also evident that most of the RHT identified 2D structures have not been identified as 3D filaments here.  

\begin{figure}[t]
\includegraphics[width=0.5\textwidth]{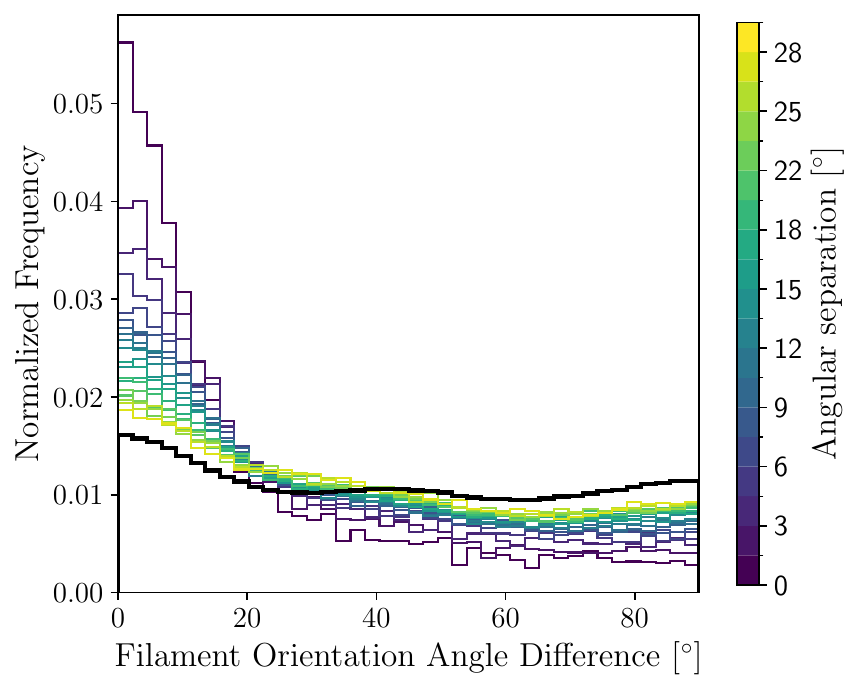}
\caption{A density histogram showing the difference in angle on the sky for filaments with different spatial separations (color bar). The black histogram shows the orientation angle difference over the full filament population. This plot shows that at smaller separations, filaments are more likely to be parallel to each other.}
\label{fig:sepang}
\end{figure}

To further assess the bundling of filaments we compared the spatial separation of filaments to the angular difference of their long axis on the sky.   We focused on the spatial separation, however Figure~\ref{sky} shows there is a clear kinematic grouping with the spatial grouping of aligned filaments. 
 Figure~\ref{fig:sepang} shows a density histogram of the distribution of filament orientation angle differences for different filament separations.  For all separation distances, the distribution of orientation angle differences is relatively flat.  The orientation angle difference gradually decreases as one goes to smaller separations between filaments. 
 When the filaments classified as local are isolated there is a sharper rise at orientation angle differences $<10$\deg~and filament separations $<20$\deg, but in general the plots are similar.



\begin{figure}[t]
\begin{center}

\includegraphics[scale=0.5]{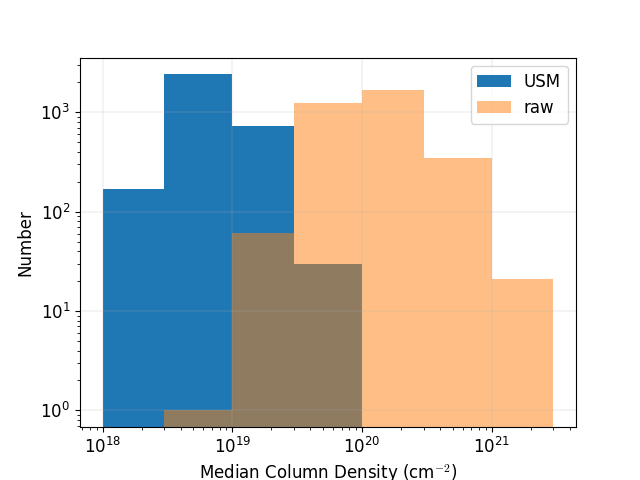}
\caption{A comparison of the median column density within the filament mask from the USM and raw data for all filaments.  There is not a large difference in values between the different filament populations.}
\label{column}    
\end{center}
\end{figure}

\begin{figure}[t]
\includegraphics[scale=0.42]{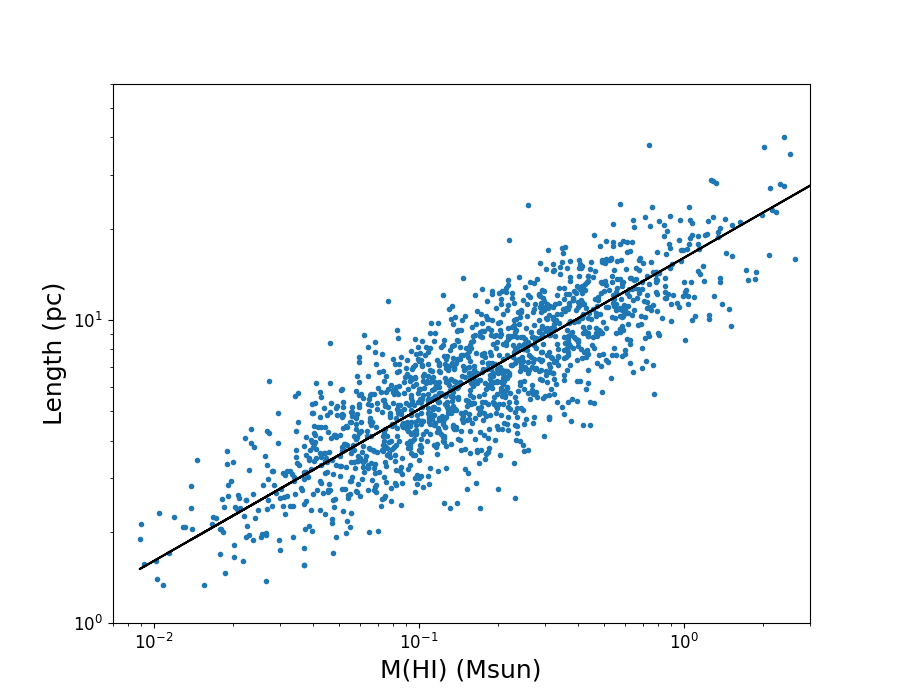}
\caption{The HI mass vs.~length relationship for the local filaments.  The line shows the L $\propto$ M$^{0.5}$ found in the filament review by \cite{hacar22}.}
\label{mhilen}
\end{figure}

\subsection{Column density and HI mass}
The distributions of filament median column densities using the raw and USM data are shown in Figure~\ref{column}.  While the USM data will underestimate the column density of the filament with the diffuse component removed, this is likely a more accurate value than the column density from the raw data, which includes all of the diffuse Galactic emission in the velocity range of the filament.  The conversion between the value from the USM to raw data can be approximated with a linear fit to the data, N$_{\rm HI(med)}$(raw) = 26.4 N$_{\rm HI(med)}$(USM) - 4.26$\times 10^{19}$ cm$^{-2}$. The median of the USM median column densities are similar for the populations, or 5.9, 6.3, and $7.7\times10^{18}$\cm~for the local, disk-halo, and other filaments, respectively.   The filaments are a small percentage of the total column density along a given line of sight.

The total column densities within the filament masks can be converted to an HI mass using the estimated distances.  The mass calculation uses the total column density listed in the table (\NH(tot) (USM)) and uses the adopted distance with the equation,
\begin{equation}
M_{HI} = N_{HI}(tot) \times m_h/m_{sun} \times l^2    
\end{equation}
where m$_h$ is the mass of a hydrogen atom and $l$ is the pixel size in cm at the distance of the object. 0.17~\Msun~ is the median mass of the local filaments from the total column density within the filament mask using the USM data.   The distribution of the local filament HI masses is shown in the length vs. HI mass diagram of Figure~\ref{mhilen}.  The line shows L $\propto$ M$^{0.5}$, which is the fit to filaments of different types found in the review by \cite{hacar22}. If the distance dependency is removed, the same relationship between angular length and total column density is found.  The local filaments are isolated here as their distance is considered more consistently reliable, but the entire population of filaments shows a similar distribution.



\begin{figure*}[t]
\begin{center}
\includegraphics[width=0.45\textwidth]{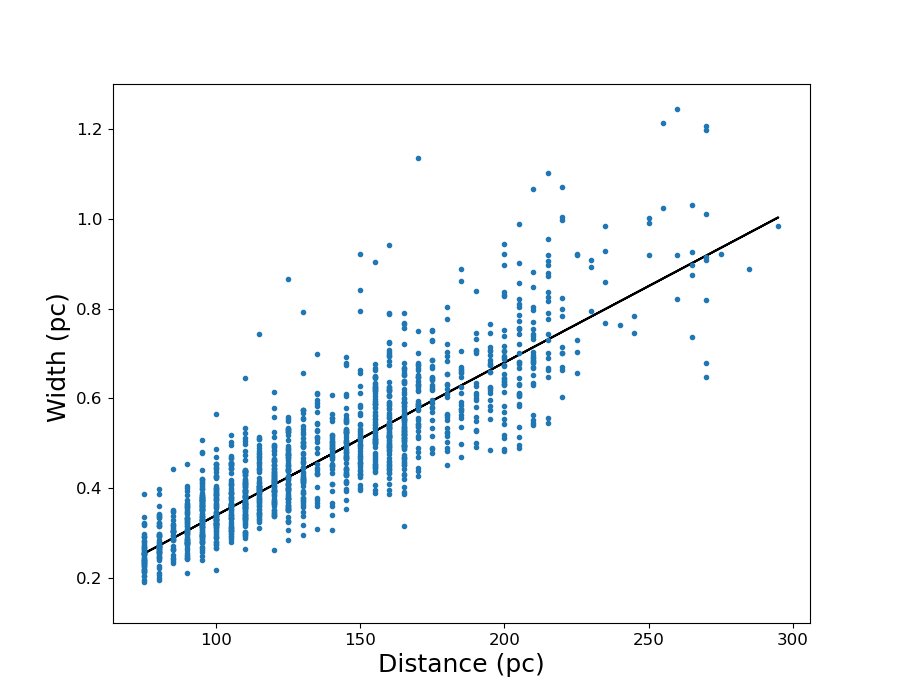}
\includegraphics[width=0.45\textwidth]{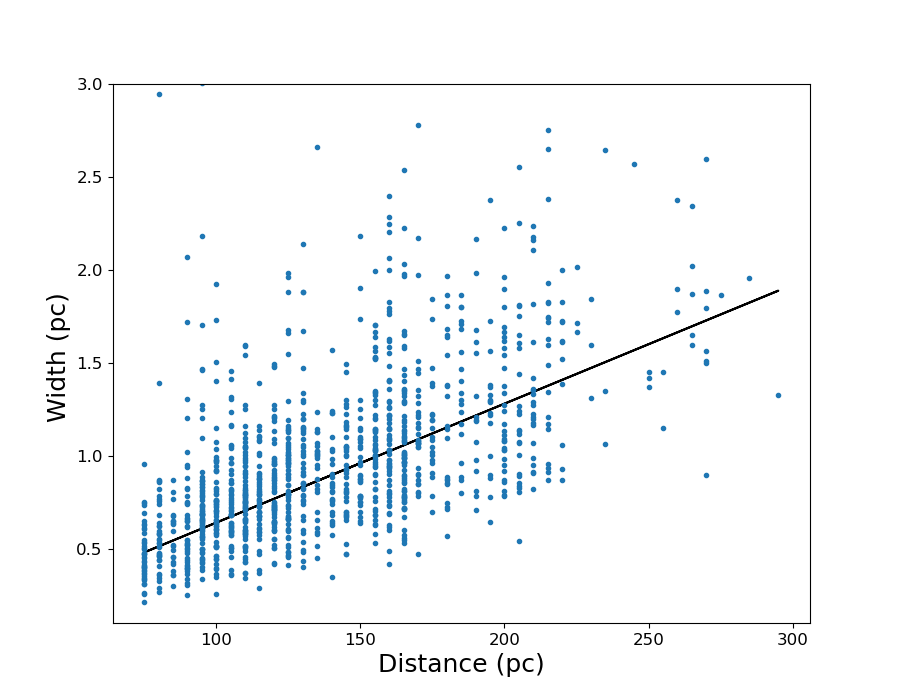}
\caption{The lower limit on the distance to the local filaments vs. the distribution of distance corrected filament widths.  These are from the {\it FilFinder} method on the USM data (left) and raw data (right). The solid line shows what the median width at a fixed distance (0.34 pc and 0.64 pc at 100 pc) would scale to at a given distances.}
\label{distwidth}
\end{center}
\end{figure*}

\begin{figure}[t]
\begin{center}
\includegraphics[scale=0.65]{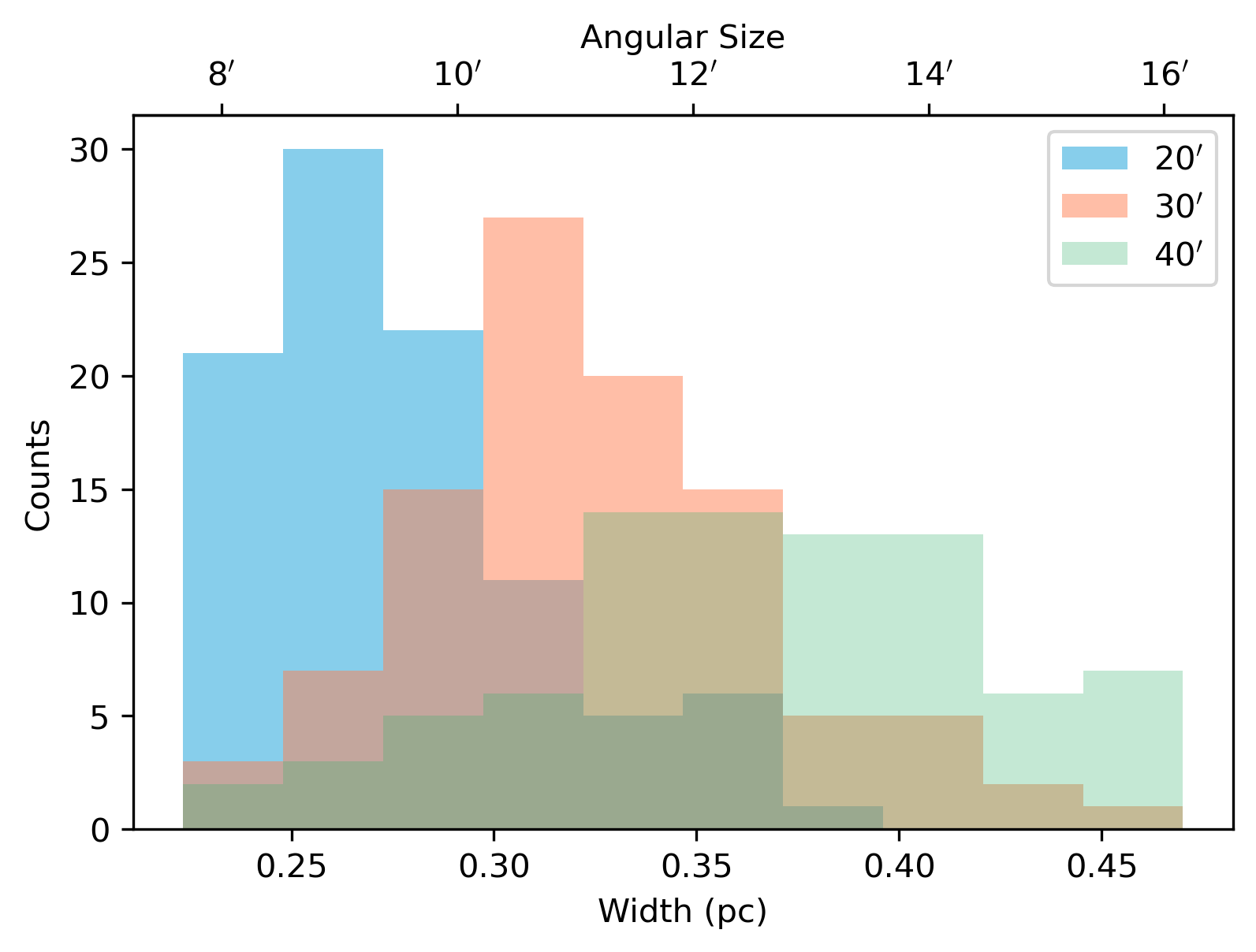}
\caption{The USM spatial width measurements for 3D HI filaments from GALFA-HI when different USM values are used (as noted in the legend).  The width on the sky is shown by the top x-axis and the physical widths at a fixed distance of 100 pc are shown on the bottom.} 
\label{widthcompare}
\end{center}
\end{figure}

\subsection{Spatial width distribution}
\label{sect:width}

Our sample of 3D HI filaments has a relatively tight distribution of spatial widths when the USM data are used.  Estimation of filament widths with the raw data is often problematic as it includes diffuse Galactic emission that can confuse the fit; on the other hand, this fit preserves any warm diffuse filament component that would typically be removed with the USM.  If the filaments are placed at a fixed distance of 100 pc, the median width of the entire population is 0.34 pc (11.7\arcmin~on the sky) using the USM data, and 0.64 pc (22.0\arcmin~on the sky) with the raw data.  The median widths remain within 0.01 pc of these values if the local or disk-halo filaments are isolated.  The spatial widths of the filaments are limited by the resolution of the data, so we primarily focus on the widths of the local filament sample.  

Figure~\ref{distwidth} shows the distance versus the physical width of the filament for the local population of filaments with the line representing how a filament with a fixed width of 0.34 pc (USM data) or 0.64 pc (raw data) scales with distance.  It is evident that the measured width scales directly with the distance to the filament using both the USM and raw data (albeit with more scatter in the latter).  
The size of the filaments is larger than the beam size (4\arcmin~or 0.1 pc at 100 pc), but the scaling with distance indicates the filaments are not well resolved. The width being larger than the beam size, but also affected by the resolution, has been found in other studies \citep{panopoulou22, andre22}.  

We tested the width measurements with various USM values in a high and low Galactic latitude region.  As shown in Figure~\ref{widthcompare}, the widths of the filaments gradually increase with larger USM values (everything else fixed).  There was no significant difference between the high and low latitude regions, so the results were merged for this plot.  The measured widths are approximately 0.05 pc (2\arcmin~at 100 pc) larger each time the USM shifts to be 10\arcmin~larger (from 20\arcmin~to 40\arcmin). The number of filaments found by {\it fil3d} remain similar with the various USM values, and, though the filaments are not precisely the same, 
the raw width distributions are close to identical.  
The USM filament widths remain consistently between approximately 0.2 - 0.5 pc with different USM values, while the raw widths remain at a median value of approximately 0.6 pc.  We also tried changing the scale width value (\S3.2) with a fixed USM value and found the width distribution of the filaments did not substantially change from a median width of 0.3 pc with the 30\arcmin~USM. 
The results are consistent with the USM kernel subtracting part of the diffuse emission that can be associated with filaments, suggesting most filaments have a outer warm and inner cold HI component.


\begin{figure}[t]
\begin{center}
\includegraphics[width=0.52\textwidth]{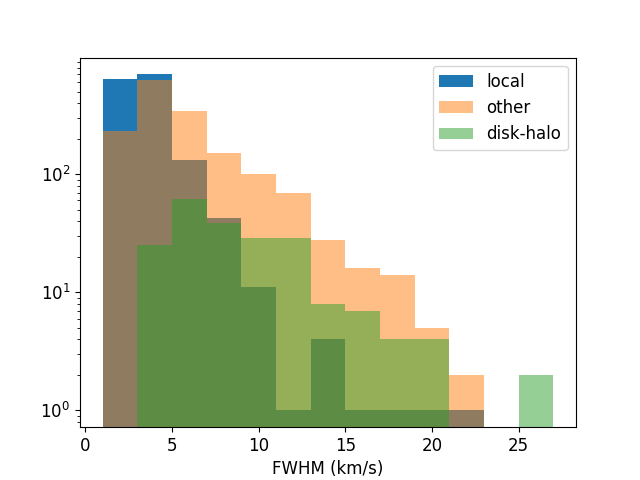}
\caption{The velocity width distribution (log scale) for the local, disk-halo and other (uncategorized) filaments in 2\kms~bins. }
\label{velwidth}
\end{center}
\end{figure}

\subsection{Velocity width distribution}

The distributions of velocity widths of the local, disk-halo and other filaments are shown in Figure~\ref{velwidth}.  As described in \S\ref{sect:cuts}, this velocity width is identified as a coherent velocity structure within the filament mask from the USM data and is not limited to the channels {\it fil3d} identifies (which are consistently at a median of 2 channels for all populations).   The level of background emission does not normally allow for a clean fit to the velocity width from the raw data. 
For the local population, the 10-90\% range of the velocity widths is 2.3-5.3\kms.  The median value of 3.2\kms~is the same value found by \cite{kim23} for local filaments using the GALFA-HI data with higher velocity resolution. The disk-halo filaments have a median velocity width of 7.7\kms~with a 10-90\% range of 5-13.8\kms, and the uncategorized filaments have a median FWHM of 4.6\kms.

 An estimate on the temperature of the filament can be calculated with T$_{\rm K}$=21.9~K(FWHM/\kms)$^{2}$.
This is usually considered an upper limit on the temperature, as a turbulent component may contribute to the linewidth.   Since the linewidth estimated here is from the USM data, the true linewidth is likely larger, and so the temperature can be considered more of an estimate than an upper limit.  The derived temperatures are consistent with the local filaments being cold neutral medium structures with a median temperature of 225 K.   The disk-halo filaments are found to be warmer structures, consistent with what is traditionally found for gas at higher latitude \citep{putman12}.  There is no strong relationship between z-height and temperature for the entire population of filaments.


\subsection{Length, volume density, and pressure}

The length, volume density, and pressure can be derived for the HI filaments, similar to how this has been done for other HI clouds \citep[e.g.,][]{McClureGriffiths:2006wx, Saul:2012gt, hsu11}.   These parameters are dependent on the measured size and distance to the filaments, and given our resolution are likely to be better determined for the local filaments.  The length distribution of the local filaments is shown in Figure~\ref{mhilen} in parsecs at the distance derived for the filaments.  The median length of the local filaments is 6.4 parsecs (166\arcmin~on the sky), with the 10-90\% range 3 - 13.4 parsecs.  The median length on the sky in arcminutes is similar for all filament populations.

 The volume density (\nH) of a filament can be calculated using the total number of hydrogen atoms divided by the volume of a cylinder (assuming this shape).   This results in a median value for the local filaments of 6.5 cm$^{-3}$, with the 10-90\% range of 2.6 - 16.4 cm$^{-3}$.  Very similar values for the volume densities are derived if the peak column density is assumed to lie at the center of the cylinder, and this value is divided by the width of the filament.  
 Though these volume densities are affected by the width scaling directly with distance, the values obtained are typical of CNM volume densities \citep[e.g.,][]{stanimirovic:2018}.  

 The thermal pressure can be estimated using, P$_{\rm th}/$k (K cm$^{-3}$)= \nH T.  Using the temperature derived from the linewidth and the volume density noted above results in a wide range of P/k values (10-90\% range of 465 - 6550 K cm$^{-3}$) for the local clouds.   This range of pressures is likely due to the widths, and therefore volume densities, not being resolved properly.  One would expect lower temperature filaments to have higher volume densities, but a similar spread of volume densities is found for the full range of filament temperatures.





\section{Discussion}

Coherent filamentary HI structures in position-velocity space are found across the sky and the catalog presented here allows for a comprehensive investigation of their properties.  Approximately 1/3 of the sky is covered by this GALFA-HI 3D filament catalog, although the velocity range of $+/- 50$\kms~does not extend across the entire disk, nor does it include high-velocity clouds.   In this section we discuss the filaments in terms of their properties as a function of Galactic environment, their overall hierarchical structure, and how they compare to other ISM structures.


\subsection{Filament Properties and Environment}

The catalog presented in this paper shows that three dimensional atomic hydrogen filaments are found throughout the Galaxy.  This was also found for 2D HI filaments \citep{peek18, soler22, kalberla16}, though these 2D objects were not cataloged as discrete structures.  The filaments surround us at low velocity ($|v_{LSR}| < 10$\kms), and this is consistent with them being structures within the atomic component of the Local Bubble.  They do not dominate the column density in a given direction and the lack of filaments in the Galactic Plane is due to the high level of confusion in this area.  As discussed further in \S\ref{sect:hier}, the filaments are found in bundles, and previous works have found that their long axes are often preferentially aligned with the Galactic magnetic field lines (\citealt{kim23, clark19}; however, see \citealt{holm24}).  The identification of distinct HI structures at 0\kms~has not traditionally been achieved and this sample represents an important method of characterizing the cold ISM in the Local Bubble. 

There are no clear trends between the local filament properties and Galactic position in terms of coordinates, z-height or distance.  On one hand, this may be expected with the likelihood of a similar Local Bubble origin; on the other hand, one may expect there to be differences given the variation in pressure as one moves out of the mid-plane.   Many of the derived properties are impacted by the limitations in the derived spatial widths.  This is one of the reasons we often focus on the local filaments when presenting filament physical properties. Future higher resolution surveys, such as GASKAP \citep{Dickey:2013wf} and WALLABY \citep{koribalski20}, will provide further information on the physical properties of 3D filaments.  


The local filaments are distinct from the rest of the 3D filaments primarily in having smaller velocity widths.  This is consistent with the local filaments being colder and in a higher pressure environment close to the Galactic Plane.  This was also found by \cite{kim23} for filaments directly overhead at 0 \kms~compared to intermediate velocity filaments.   Another consistency with \cite{kim23} is that the shorter filaments do not have significantly larger velocity widths than the longer filaments.   If velocity gradients along the length of the filaments were present, we may expect shorter filaments that represent filaments that are inclined in the direction of our line of sight, to have larger velocity widths.   The longer filaments do have slightly narrower velocity widths, but are primarily consistent with the overall distribution found in Figure~\ref{velwidth}.

The filaments at higher velocities are more sparsely distributed across the sky.  
These filaments largely represent more distant structures in the rotating Galactic disk or disk-halo gas beyond the Local Bubble.  Some of the filaments can be identified as parts of known intermediate velocity cloud complexes.  The negative velocity filaments near the Galactic Plane in the top panel of Figure~\ref{sky} may be related to what is known as the Outer Arm complex and the positive velocity filaments to complex gp \citep{wakker04}.
In the middle panel of Figure~\ref{sky}, the negative velocity filaments between RA = 160 to 200\deg~at high positive Galactic latitude can be associated with the IV Arch and IV Spur at approximately 1 kpc \citep{kim23, wakker04}.  The positive velocity filaments at around b=30\deg~in this panel are likely related to the positive velocity compact Q3 clouds identified by \citet{saul14}.  While the split of positive and negative velocity filaments are expected in the plane of the Galaxy in the last panel, 
there are additional negative velocity filaments that are likely related to the intermediate velocity AC shell.   Many of these named complexes have known distances and are between 0.5-2 kpc in z-height, which is similar to our derived z-heights.  The lengths at 1 kpc extend to 100 pc scales and the masses to 10's M$_\odot$.  The widths will likely be significantly unresolved at these distances.  The disk-halo filaments are warmer in nature (Figure~\ref{velwidth}) and it is not yet easily possible to determine if they are aligned with the magnetic field in that region \citep[e.g.,][]{kim23,holm24}.  Future polarization measurements of stars with known distances will help disentangle the magnetic field uncertainty \citep[e.g.,][]{panopoulou22}.  The existence of these filaments may be due to linear structures produced by ram pressure forces and/or cooling as they move through the inner Galactic halo \citep[e.g.,][]{abruzzo24,abruzzo22}.

 This catalog shows 3D HI filaments are ubiquitous ISM structures in the shell of neutral gas surrounding the Sun.   It also shows that 3D filaments are commonly found in other HI structures throughout the Milky Way and that methods can be developed to quantify the properties of these diffuse structures.  Simulations of molecular clouds have found the nature of the molecular filaments can be affected by line of sight confusion \citep{zamora17,beaumont13}.  This should be investigated with simulations of the HI filaments. 

\subsection{Hierarchical Structure}
\label{sect:hier}

The 3D filaments show evidence of a hierarchical structure with their relationship between length and mass and the bundling of filaments with similar velocities and alignments on the sky.  This could suggest there are larger linear HI structures breaking down into smaller structures, or that the smaller structures are building up into larger structures.  This relationship between length and mass has been found for filaments detected at many different wavelengths throughout our Galaxy and is consistent with turbulence playing an important role in their formation \citep{Hacar:2013uh,hacar22,kalberla25, saury14}. 

An overall hierarchical structure for the filaments should be evident in comparing surveys of different resolutions and filaments at different distances. 
 As long as the sensitivity is sufficient, more filaments should be detected in higher resolution surveys or at the closest distances.   The initial work on the HI filaments in large surveys by \cite{clark14} compared filamentary structure in the GALFA-HI survey compared to the Parkes Galactic All-Sky Survey with a factor of 4 difference in resolution.  Indeed, more filaments were detected in the GALFA-HI data; however, this would need further analysis to confirm it is showing hierarchy of structure and not just resolution effects.  
This catalog provides an opportunity to investigate whether there is a decrease in the number of filaments with increasing distance.  Though the distances used here are only estimates, any variation on a plot of the distribution of number of filaments with distance shows a decrease in number with distance (i.e., only the local filaments, only high Galactic latitude, only kinematic distances, etc.).  Again, future work can consider the sensitivity to filaments at various distances to further support that the decreasing number with distance supports a hierarchical model.  

The driving scale of the filamentary structure can potentially be inferred from the size of the bundles of aligned filaments on the sky.  There is an increase in the number of filaments with orientation angle differences $<10$\deg~with decreasing separation on the sky, as shown in Figure~\ref{fig:sepang}.   It is difficult to identify a preferred scale from this figure, but the most distinct increase for small orientation angular differences is at $<10$\deg~separation between filaments.  At 100-200 parsecs this separation corresponds to roughly 15-30 parsecs, which is consistent with the expected driving scale for stellar feedback \citep{colman22}. Turbulence, combined with other mechanisms such as thermal instability \citep[e.g.,][]{audit05, saury14} and the strength of the shear strain relative to the magnetic field \citep{Inoue:2016wb}, could produce these cold filaments within the Local Bubble.

\subsection{Comparison to other ISM structures}


We compare the properties of the 3D HI filaments to other filaments and HI features with various morphologies to gain insight into their relationship to the broader ISM.  There are a wealth of references on atomic, dusty, and/or molecular filaments in the Galactic ISM and we focus on the papers with relevant comparison points on physical properties here.

Previous work in cataloging HI ISM structures has focused on features beyond 0\kms~given the complexity of extracting discrete HI features when bright diffuse emission fills the field.  We can most directly compare our HI filaments to other GALFA-HI cloud catalogs, e.g., the compact HI clouds at a range of velocities beyond 0 \kms~by \cite{Saul:2012gt} and the high velocity clouds of Complex C and the Magellanic Stream catalogued by \cite{hsu11}.   The clouds in both catalogs have larger velocity widths overall, consistent with them being warmer structures at higher latitude and also not having the diffuse component subtracted.  There is a population of cold compact clouds in Saul et al.~scattered across the GALFA-HI sky area with a median velocity of -38\kms~that have similar velocity widths and column densities to our disk-halo filaments.  Though the Saul et al. compact clouds were required to be free of connecting emission, it could be that the connecting emission was not always detectable.  It could also represent the range of structures that can form in a turbulent medium \citep[e.g.][]{abruzzo24}. 

The most direct comparison to these 3D filaments is to other HI filaments.   The GALFA-HI 3D filaments studied by \cite{kim23} have very similar properties (e.g., FWHM, N(HI), length) to the filaments cataloged here.   This is expected, but still reassuring that the 0.75 vs.~0.18\kms~channels does not significantly affect the derived properties.  It is likely the majority of the filaments in the catalog are similar to the Kim et al.~sample in that they do not show velocity gradients along their length, but the higher velocity resolution data is better suited for studying this.   

We have not examined the alignment of the filaments with the magnetic field in this paper.  The GALFA-HI 3D filaments at high-latitude are found to be aligned with the magnetic field and those closer to the disk studied thus far are not well aligned \citep{kim23,holm24, soler22}.   We expect the majority of the aligned bundles to be aligned with the Galactic magnetic field given the similarity of some of the bundles to the 2D RHT identified GALFA-HI filament distributions \citep{clark14,peek18} and other work which finds filamentary HI structures aligned with the Galactic magnetic field \citep{kalberla21}.

 The widths of filaments have captured attention, but as found with the molecular filaments \citep{panopoulou22, andre22}, measuring precise values is elusive.   We can confidently say that the filaments are sub-parsec width structures with a CNM component that is unresolved.  This is supported in that there is no clear relation between the width of the filament and the column density for the local population of filaments. The fact that the filament width is sensitive to the USM size applied indicates the filaments do not have a sharp drop-off from the general Galactic background emission.  This is consistent with a shallow profile that would be expected for magnetized filaments \citep[e.g.][]{fiege00}, though this theoretical work focused on molecular structures.  Despite some variation with the background subtraction, the most local filaments are consistent with being $<0.6$~pc in width.  Within the limitations of the data, the HI filament widths derived are of similar magnitude to those of denser filaments (0.1 - 1 pc; \citet{hacar22}).

\section{Conclusions}
\label{conc}

Neutral hydrogen filaments are found throughout the ISM. In this paper, we present a 3D filament finder to identify HI filaments that are continuous in position-position-velocity space across the 13,000 deg$^{2}$ of the GALFA-HI survey (4\arcmin~resolution, 0.75\kms~channels; \cite{peek17}), and derive their physical properties.
Some of the primary takeaways from the paper can be summarized as follows:
\begin{itemize}
\item  We present a method to catalog filaments that are continuous in position-position-velocity space called {\it fil3d}, which builds on the 2D filament finder {\it FilFinder} \citep{Koch:2015dc}.  The {\it fil3d} code is distributed via GitHub at \url{https://github.com/a-dykim/fil3d}. 
The algorithm is used to identify 3333 3D filaments with $|v_{\rm LSR}| < 50$\kms.  The positions, velocities, velocity widths, lengths, widths, orientations, and column densities of the filaments are cataloged.

\item   Based on a filament's position and velocity, we are able to estimate the distance to approximately half of the filaments.  We classify 1542 filaments as local structures along the neutral shell of the Local Bubble at distances $<200$~pc, and 209 filaments as disk-halo, at z-heights of $\sim1$ kpc.  The local filaments have a 10-90\% range of velocity widths of 2.3-5.3\kms, representative of the CNM; the disk-halo velocity widths are twice as large.

\item  The 3D filaments are found across the sky in aligned bundles and the local filaments follow the L $\propto$ M$^{0.5}$ relationship found for filaments at other wavelengths.  These filament properties are evidence for their hierarchical structure and is consistent with turbulence playing an important role in their formation.

\item  Using the data with a USM applied, the local filaments have a typical width of 0.34 pc at 100 pc or $\sim12$\arcmin.  For those local filaments that can have their width fit from the raw data, they have a median width fit of 0.64 pc at 100 pc.  This is consistent with the USM partially removing a more extended diffuse component.   The widths remain similar at all filament distances, indicating the resolution limitations in our study.

\end{itemize}

The tools presented here can be used on new high resolution HI surveys of our Galaxy (e.g., GASKAP \citep{Dickey:2012wf}) and of other galaxies (e.g., LGLBS (Koch et al. 2025)).  The use on other galaxies can determine if atomic HI filaments are a generic feature in a galaxy's ISM \citep{ma23}.   These tools can also be used on H$\alpha$ surveys in the future to study the ionized counterparts of HI filaments \citep[e.g.,][]{aftab24,melso24}.

{\it Acknowledgement.} 
The authors thank the GALFA-HI and ALFALFA teams that made the data used here possible, and Garrison Groger for help with the initial code development.  We thank the referee for useful feedback on the paper. This project was partially supported by Cottrell SEED Award \#CS-SEED-2023-009 from the Research Corporation for Science Advancement.
This project makes use of astropy \citep{astropy1, astropy2, astropy3}, FilFinder\citep{Koch:2015dc}, healpy \citep{healpy1, healpy2}, numpy and scipy \citep{scipy}, and matplotlib \citep{matplotlib}.

\section*{Data availability}
This publication utilizes data from Galactic ALFA HI (GALFA-HI) survey data set obtained with the Arecibo L-band Feed Array (ALFA) on the Arecibo 305 m telescope (available at https://purcell.ssl.berkeley.edu/).

\bibliography{refs}{}
\bibliographystyle{aasjournal}

\end{document}